\documentclass[11pt,a4paper,cite]{article}
\usepackage[english]{babel}
\usepackage[T1]{fontenc}
\usepackage[latin1]{inputenc}
\usepackage{epsfig}
\usepackage{graphics}
\usepackage{cite}
\usepackage{amsmath}
\usepackage{amsfonts}
\usepackage{amssymb}
\usepackage{amsthm}
\usepackage[makeroom]{cancel}
\usepackage{color}
\usepackage{bm}
\usepackage{datetime}
\usepackage{xcolor}

\newcommand{\beq}{\begin{equation}}
\newcommand{\eeq}{\end{equation}}

\newcommand\blfootnote[1]{%
	\begingroup
	\renewcommand\thefootnote{}\footnote{#1}%
	\addtocounter{footnote}{-1}%
	\endgroup
}

\newdateformat{monthyeardate}{%
	\monthname[\THEMONTH], \THEYEAR}

\title{\Large \bf A class of solitons in Maxwell-scalar \\ and Einstein-Maxwell-scalar models}
\author{Carlos A. R. Herdeiro$^{\ddagger}$, Jo\~{a}o M. S. Oliveira$^{\dagger}$, Eugen Radu$^{\star}$}
\date{%
	{\small Departamento de Matem\'atica da Universidade de Aveiro and CIDMA,} \\ {\small Campus de Santiago, 3810-183 Aveiro, Portugal}\\
 \vspace{0.3cm}
	\monthyeardate\today
}

\begin{document}
\maketitle
\blfootnote{$\ddagger$ herdeiro@ua.pt}
\blfootnote{$\dagger$ jmiguel.oliveira@ua.pt}
\blfootnote{$\star$ eugen.radu@ua.pt}

\begin{abstract} 
Recently, no-go theorems for the existence of solitonic solutions in Einstein-Maxwell-scalar (EMS) models have been established~\cite{Herdeiro:2019oqp}. Here we discuss how these theorems can be circumvented by a specific class of non-minimal coupling functions between a real, canonical scalar field and the electromagnetic field. When the non-minimal coupling function diverges in a specific way near the location of a point charge, it regularises all physical quantities yielding an everywhere regular, localised lump of energy. Such solutions are possible even in flat spacetime Maxwell-scalar models, wherein the model is fully integrable in the spherical sector, and exact solutions can be obtained, yielding an explicit mechanism to de-singularise the Coulomb field. Considering their gravitational backreaction, the corresponding (numerical) EMS solitons provide a simple example of self-gravitating, localised energy lumps. 
\end{abstract}

\newpage



\section{Introduction}
Einstein-Maxwell-scalar (EMS) models, described by the action
\begin{equation}
\mathcal{S}=\frac{1}{4\pi}\int d^4x \sqrt{-g} \left(\frac{R}{4}-\frac{f(\phi)}{4}F_{\mu\nu}F^{\mu\nu}-\frac{1}{2}\partial_\mu\phi \partial^\mu \phi\right) \ ,
\label{actiontotal}
\end{equation}
wherein the real scalar field $\phi$ has a canonical kinetic term and it is non-minimally coupled to the Maxwell field strength $F$, via some function $f(\phi$), emerge naturally in physics ($R$ is the Ricci scalar). Well known contexts are Kaluza-Klein models~\cite{Kaluza:1921tu,Klein:1926tv,Appelquist:1987nr} and supergravity/string theory~\cite{VanNieuwenhuizen:1981ae}. In these cases the non-minimal coupling is provided by an exponential function of the sort $f(\phi)\sim e^{-\alpha \phi}$, where $\alpha$ is a constant. But more general classes of coupling functions have been considered, for instance, in the context of cosmology~\cite{Martin:2007ue,Maleknejad:2012fw}. 

Another interesting class of coupling functions emerged recently in the quest for models that deviate from general relativity only in the strong gravity regime. If $f(\phi)$ obeys some simple properties, the EMS model accommodates the phenomenon of spontaneous scalarisation of asymptotically flat charged black holes~\cite{Herdeiro:2018wub}. This means that even though the electrovacuum Reissner-Nordstr\"om (RN) black hole is a solution of the EMS model, for sufficiently high charge to mass ratio this black hole becomes unstable: it becomes energetically favourable for the RN black hole to scalarise. A new family of scalarised black holes bifurcates from the RN family, which contains the end states of this dynamical scalarisation mechanism - see also~\cite{Fernandes:2019rez,Astefanesei:2019pfq,Myung:2018vug,Myung:2018jvi,Myung:2019oua,Konoplya:2019goy,Fernandes:2019kmh}.

The existence of such a class of EMS models, that allow spontaneous scalarisation of charged black holes, and accommodates two distinct families of black holes, the bald RN and the ``hairy" scalarised BHs, raised the question if solitons can also be found in EMS models, as this is often the case in theories that accommodate both bald and hairy BHs. Building on this question, in~\cite{Herdeiro:2019oqp} a set of theorems were established showing that, under various assumptions, no such solitonic solutions exist, similarly to the case of vacuum and electrovacuum. The purpose of this paper is to show that dropping one of the assumptions considered in~\cite{Herdeiro:2019oqp}, namely that the coupling function is everywhere finite, it becomes possible to circumvent the aforementioned theorems and obtain solitonic solutions, including in the flat spacetime (Maxwell-scalar theory) limit. In the latter case, moreover, the solitons can be, in some examples, obtained in closed form. They yield pedagogical illustrations of how new physics could de-singularise the Coulomb solution at the level of an effective field theory. In the self-gravitating case, the solutions are obtained numerically, although we cannot exclude that some carefully designed coupling functions exist where they will have a closed analytic form. 

This paper is organised as follows. In Section~\ref{sec2} we discuss the flat spacetime analysis, $i.e.$ the Maxwell-scalar model in Minkowski spacetime. We show that the model is integrable and how the Coulomb singularity can be regularised by a divergent coupling in a class of explicit examples.  In Section~\ref{sec3} we consider the self-gravitating solitons, corresponding to the solutions discussed in flat spacetime in Section~\ref{sec2}. Conclusions are presented in Section~\ref{sec4}.

\section{Flat spacetime Maxwell-scalar models}
\label{sec2}

\subsection{A physical motivation}
An awkward feature of classical electromagnetism is that the energy $E$ of the Coulomb field of a point charge $Q$ is divergent:
\begin{equation}
E\sim \int_{0}^{+\infty}\frac{Q^2}{r^2}dr=\infty \ .
\end{equation}
Quantum considerations naturally introduce an ultraviolet cut-off to the validity of the classical Coulomb solution, regularising this integral. Quantum Electrodynamics (QED), however, is itself incomplete as a quantum field theory, due to the Landau pole~\cite{Peskin:1995ev}. But it yields the important lesson that the coupling constant $g$, which determines the strength of the electromagnetic interaction in the Maxwell Lagrangian
\begin{equation}
\mathcal{L}=-\frac{1}{4g^2}F_{\mu\nu}F^{\mu\nu} \ ,
\end{equation} 
runs with the energy scale. 

Whatever fundamental theory turns out to complete QED, it may admit a covariant effective field theory description that captures the dynamics of the coupling. Then, $g$ would emerge as a spacetime function with some dynamics. In a simple model, $g$ would be a real scalar field with a standard kinetic term. Allowing a more general dynamics, one takes $g$ as being an arbitrary function of the scalar field, keeping the latter with a standard kinetic term. This suggests considering the \textit{naive} covariant effective field theory
\begin{equation}
\mathcal{S}=\frac{1}{4\pi}\int d^4x \sqrt{-g} \left(-\frac{f(\phi)}{4}F_{\mu\nu}F^{\mu\nu}-\frac{1}{2}\partial_\mu\phi \partial^\mu \phi\right) \ ,
\label{action}
\end{equation}
where $\phi$ is a real scalar field,  $F=dA$ is the covariant description of the electromagnetic dynamics and the background is Minkowski spacetime. The function $f(\phi)$ specifies the dynamics of the gauge coupling. This model ignores higher order corrections in $F$, so it is certainly incomplete. Nonetheless one may take the aforementioned reasoning as a motivation to consider this class of simple models. Can the Coulomb field of a point charge be de-singularised in this context?

\subsection{An integrable model }
The naive model~\eqref{action}, which is the decoupling limit of the EMS model~\eqref{actiontotal} wherein back reaction is neglected, is integrable in the spherical sector. Taking the following ansatz for the fields in spherical coordinates in Minkowski spacetime $(t,r,\theta,\varphi)$:
\begin{equation}
\phi=\phi(r) \ , \qquad A=V(r)dt \ ,
\label{ansatz}
\end{equation}
the Maxwell equations yield a first integral:
\begin{equation}
 V(r)=\int \frac{Q}{r^2f(\phi)}  dr \ ,
\label{maxint}
\end{equation}
where $Q$ is interpreted as the \textit{electric charge}. Using this first integral, the Klein-Gordon equation reads
\begin{equation}
r^2\frac{d}{dr}\left(r^2 \frac{d\phi}{dr}\right)-\frac{Q^2}{2}\frac{d}{d\phi}\left(\frac{1}{f(\phi)}\right) = 0  \ ,
\end{equation}
which, introducing the coordinate $x\equiv 1/r$, yields another first integral
\begin{equation}
\left(\frac{d\phi}{dx}\right)^2-\frac{Q^2}{f(\phi)}=\mathcal{E} \ .
\label{eq2}
\end{equation}
It is a simple application of the virial theorem, or a Derrick-type scaling theorem~\cite{Derrick:1964ww}, to show that solutions must have $\mathcal{E} =0$. For instance, this can be seen from the condition~\cite{Gibbons:1990um}
\begin{equation}
\int d^3x T_{ij}=0 \ ,
\label{virialg}
\end{equation}
that holds for time-independent, finite energy field configuration in Minkowski spacetime, where $i,j$ are spatial indices in Cartesian coordinates. Relation~\eqref{virialg} is a simple consequence of energy-momentum conservation and can be interpreted as the balancing of the total stresses in an extended object. There are regions where matter is in tension and regions where it is in compression, for any static balanced soliton. Thus, the problem of finding solutions is reduced to solving, from~\eqref{eq2},
\begin{equation}
x(\phi)=\frac{1}{Q} \int \sqrt{f(\phi)}d\phi \ ,
\label{hj}
\end{equation}
and then inverting $x(\phi)\rightarrow \phi(x)\rightarrow \phi(r)$. Fixing the coupling function $f(\phi)$ one can obtain $\phi(r)$ and, from~\eqref{maxint}, the electrostatic potential,  both as line integrals. Due to the two first integrals the system is fully integrable.

\subsection{Everywhere regular solutions}
\label{regsol}

To assess if the solutions have finite energy one must consider the energy-momentum of the model, 
\begin{equation}
4\pi T_{\mu\nu}=f(\phi)\left(F_{\mu\alpha}F_\nu^{\ \alpha}-\frac{1}{4}g_{\mu\nu}F_{\alpha\beta}F^{\alpha \beta}\right)+\partial_\mu\phi \partial_\nu\phi -\frac{1}{2}g_{\mu\nu}\partial_\alpha \phi \partial ^\alpha \phi \ .
\end{equation}
This yields the energy density $\rho$, after using~\eqref{eq2}:
\begin{equation}
\rho=T_{00}= \frac{Q^2}{4\pi r^4f(\phi)}  \  \ .
\label{edensity}
\end{equation}
and the total energy, $E$,  obtained by integrating the energy density on a spacelike slice $\Sigma$
\begin{equation}
E=\int_{\Sigma}  \rho \, d^3x= \int_{0}^{+\infty} \frac{Q^2}{r^2f(\phi)} dr\ .
\label{totale}
\end{equation} 

In order to obtain regular solutions at the origin we assume the scalar field admits a power series expansion near the origin:
\begin{equation}
\phi= \phi_0+\sum_{p=N}\phi_pr^p \ ,
\label{psphi}
\end{equation}
We do not constrain the constant coefficient $\phi_0$, which may or may not vanish. Apart from $\phi_0$, let $\phi_N$, where $N\in \mathbb{N}\geqslant1$ be the first non-vanishing coefficient in this expansion. Then,  from~\eqref{eq2},
\begin{equation}
(-Nr^{N+1}\phi_N+\dots)^2=\frac{Q^2}{f(\phi)} \ .
\end{equation}
Thus, as $r\rightarrow 0$, 
\begin{equation}
f(\phi)\sim \frac{Q^2}{N^2\phi_N^2}\, \frac{1}{r^{2N+2}} \ .
\label{cdiv}
\end{equation}
Regularity of the scalar field at the origin then~\textit{requires} the coupling to diverge as $\sim 1/r^{2N+2}$. From~\eqref{edensity} this implies that the energy density is finite therein and from~\eqref{maxint}, 
\begin{equation}
V(r)=V(0)+\frac{N^2\phi_N^2}{(2N+1)Q}\, r^{2N+1}+\dots \ ,
\end{equation}
close to the origin. Thus, all physical quantities are finite close to the origin, for this class of behaviours of the coupling.

\subsection{A class of examples}
There is still, of course, some freedom in choosing the coupling function, within the class with the correct divergent behaviour at the origin. Let us consider examples.

\subsubsection{A simple coupling yielding regular solutions}
\label{sec241}

As an explicit example, consider 
\begin{equation}
f(\phi)=\frac{1}{(1-\alpha \phi)^4} \ ,
\label{coupling1}
\end{equation}
where $\alpha$ is a non-zero constant.  Then~\eqref{hj} immediately yields, taking the integration constant such that $\phi\rightarrow 0$ as $r\rightarrow \infty$:
\begin{equation}
\phi(r)=\frac{Q}{Q\alpha+r} \ .
\label{phi1}
\end{equation}
One observes that $\phi(r)$ is regular and smooth as $r\rightarrow 0$, $\phi(r)\simeq1/\alpha -r/(Q\alpha^2)$; thus we expect, from~\eqref{cdiv}, that the coupling to diverge as $1/r^4$. Asymptotically , on the other hand, $\phi(r)\simeq Q/r$. Thus the scalar ``charge" coincides with the electric charge.  Plugging~\eqref{phi1} into~\eqref{coupling1} yields:
\begin{equation}
f(r)=\left(1+\frac{\alpha Q}{r}\right)^4 \ .
\label{coupling1r}
\end{equation}
The coupling diverges as $1/r^4$ at the origin, as anticipated. This divergence precisely cancels the divergence of the Maxwell field at the origin, $cf.$~\eqref{edensity}, making it finite and non-zero. In fact, the energy  density, from~\eqref{edensity}, is
\begin{equation}
\rho=\frac{Q^2}{4\pi (Q\alpha+r)^4} \ .
\label{edensity1}
\end{equation}
It follows that the total energy~\eqref{totale} is
\begin{equation}
E=\frac{Q}{3\alpha} \ .
\label{totale1}
\end{equation}

Now, using~\eqref{maxint} we obtain for the electrostatic potential:
\begin{equation}
V(r)=-\frac{rQ}{(Q\alpha+r)^2}-\frac{\alpha^2Q^3}{3(Q\alpha+r)^3} \ .
\label{v1}
\end{equation}
All the quantities~\eqref{edensity1}, \eqref{totale1}, \eqref{v1} manifestly reduce to the usual Coulombic ones upon taking $\alpha\rightarrow 0$. In such case~\eqref{phi1} reduces to the profile of a scalar charge $Q$ at the origin. The expressions make manifest how $\alpha$ regularises the solution.

\subsubsection{A family of couplings yielding regular solutions}

As further examples, with slightly different features, we generalise the coupling~\eqref{coupling1} as 
\beq
f(\phi) = \frac{1}{(1-\alpha\phi)^n} \ ,
\label{cfgen}
\eeq
where $n$ is an integer. Using this coupling, equation \eqref{hj} gives
\beq\label{integ}
\frac{1}{r} = \frac{1}{Q}\int (1-\alpha\phi)^{-n/2} d\phi \ ,
\eeq
which has a different indefinite integral for $n\neq 2$ and $n=2$.

For $n\neq 2$, imposing $\phi(r\rightarrow\infty) = 0$ to fix the integration constant, one obtains 
\beq\label{scalarfield}
\phi(r) = \frac{1}{\alpha} - \frac{1}{\alpha}\left[1+\frac{\alpha Q(n-2)}{2r}\right]^{\frac{2}{2-n}} \ ,
\eeq
which reduces to~\eqref{coupling1} for $n=4$. For regular solutions at the origin we require $\lim_{r\rightarrow 0} \phi$ to be finite. This  implies $n>2$, in which case 
\begin{equation}
\lim_{r\rightarrow 0}\phi(r) =\frac{1}{\alpha}-\frac{1}{\alpha}\left(\frac{2r}{\alpha Q(n-2)}\right)^{\frac{2}{n-2}} \ ,
\end{equation}
which is finite, as required. For $n=3$ we see that the second term goes as $r^2$; but for $n>4$, the second term has a non-integer power. In the former case we anticipate, from~\eqref{cdiv}, that  the coupling diverges as $1/r^6$. In the latter case, $\phi$ is not analytic at the origin. It will, nonetheless yield a regular solution, when analysing the usual physical quantities. 

The coupling $f(\phi)$ as a function of $r$ then reads:
\beq\label{ffunction}
f(r) = \left[1+\frac{\alpha Q}{2r}(n-2)\right]^{\frac{2n}{n-2}} \ ,
\eeq
which diverges as $\sim1/r^{\frac{2n}{n-2}}$ at the origin, for $n>2$, but respects the condition $\lim_{r\rightarrow\infty} f(r)= 1$. We confirm, in particular, the $1/r^6$ divergence, for $n=3$ and a divergence with (generically) a non-integer inverse power for $n\geqslant 5$. The electric field $E_\mu = -\partial_\mu V(r)$ has only one non-zero component which reads, from~\eqref{maxint} 
\beq \label{Efield}
E_r(r) = -\frac{Q}{r^2f} = -\frac{Q}{r^2}\left[1+\frac{\alpha Q}{2r}(n-2)\right]^{-\frac{2n}{n-2}} \ ,
\eeq
which behaves as $r^\frac{4}{n-2}$ near the origin, and it is thus regular for $n>2$.

The total energy now reads
\begin{align}\label{Energy}
E =\frac{2 Q}{(n+2)\alpha} \ .
\end{align}
Thus, the family of cases with $n>2$ illustrate how regular solutions can be obtained, with a different analytic behaviour  of the scalar field near the origin (the cases $n=3$ and $n=4$) and non-analytic behaviour ($n>4$).

With $n=2$, following a similar reasoning one obtains
\beq
\phi(r) = \frac{1}{\alpha}\big(1-e^{-\alpha Q/r}\big) \ ,
\eeq
which is a regular solution at $r=0$ with $\lim_{r\rightarrow 0}\phi(r)=1/\alpha$.
The coupling function $f(\phi)$ becomes
\beq
f(r) = e^{2\alpha Q/r} \ ,
\eeq
which, as before, also diverges at $r=0$ but respects $\lim_{r\rightarrow \infty} f(r) = 1$. Observe, however, it does not diverge as an inverse power of $r$, which was the conclusion in Section~\ref{regsol}. This is because,  again, $\phi$ in this case does not admit a power series expansion near the origin. This illustrates yet a different example of divergent coupling that yields regular solutions. 

The electric field is now
\beq
E_r(r) = -\frac{Q}{r^2 f} = -\frac{Q}{r^2}e^{-2\alpha Q/r} \ ,
\eeq
and the total energy is
\begin{align}
E =\frac{ Q}{2\alpha}  \ .
\label{enen2}
\end{align}
In these considerations $\alpha Q$ was assumed to be positive. Otherwise the total energy \eqref{Energy}-\eqref{enen2} would be negative, which would violate  the weak energy condition. Interestingly enough, despite the seemingly different solution for $n=2$, the total energy $E$ is a smooth function of the power $n$, as~\eqref{enen2} coincides with setting $n=2$ in \eqref{Energy}.

\subsection{Dilatonic coupling: a spherically symmetric solution}
As mentioned in the Introduction, a dilatonic coupling 
 \begin{eqnarray}  
 f(\phi)=e^{-\alpha \phi} \ ,
 \label{dila}
\end{eqnarray}
where $\alpha$ is a constant, emerges in relevant scenarios.  Let us thus briefly mention the existence of a  spherically symmetric, exact solution for this coupling. 

Considering~\eqref{dila} in~\eqref{hj}, and taking the integration constant so that  the scalar field vanishes asymptotically we immediately get
 \begin{eqnarray}  
 \phi=-\frac{2}{\alpha}\log\left[1+\frac{\alpha Q}{2r}\right]\ .
\end{eqnarray}
Thus, the coupling, as a function of $r$ is
\beq
 f(\phi)=e^{-\alpha \phi}=\left[1+\frac{\alpha Q}{2r}\right]^2 \ .
\eeq
Thus, the coupling diverges at the origin and, if $\alpha Q>0$ it is regular elsewhere. Moreover, using now~\eqref{maxint} we get
\beq
  \qquad V(r)=-\frac{2Q}{\alpha Q+2 r} \ .
\eeq
One finds the following small-$r$ expansion of the solution
 \begin{eqnarray}  
\phi(r)=\frac{2}{\alpha}\left(\log r-\log\frac{\alpha Q}{2} \right)+\mathcal{O}(r) \ , \qquad
V(r)=-\frac{2}{\alpha}+\frac{4r}{\alpha^2 Q}+\mathcal{O}(r^2) \ ;
\end{eqnarray}
thus, the scalar field diverges at the origin. Asymptotically, on the other hand, both fields are well behaved
 \begin{eqnarray}  
 \phi(r)=-\frac{Q}{r}+\frac{\alpha}{4}\frac{Q^2}{r^2}+\mathcal{O}\left(\frac{1}{r^3}\right) \ , \qquad 
  V(r)=-\frac{Q}{r}+\frac{\alpha}{2}\frac{Q^2}{r^2}+\mathcal{O}\left(\frac{1}{r^3}\right) \ .
\end{eqnarray}
	The energy density of this solution diverges at the origin:
	 \begin{eqnarray}  
\rho=-T_t^t=\frac{Q^2}{\pi r^2(\alpha Q+2r)^2} \ ;
\end{eqnarray}
the total mass, however,  is finite
 \begin{eqnarray}  
M= 4\pi \int_0^\infty dr r^2\rho=\frac{2 Q}{\alpha} \ .
\end{eqnarray}
This solution is interesting in that it shows a divergent coupling can source a finite mass configuration which, nonetheless, is not fully regular, as the scalar field and the energy density diverge at the origin.

\section{The gravitating solitons}
\label{sec3}
The above flat spacetime solutions can be made to self-gravitate by coupling~\eqref{action} to Einstein's general relativity. For the case of the regular solutions described in the previous section, this yields, perhaps, the simplest models of charged soliton. 

One now considers the EMS model~\eqref{actiontotal}, where  $c=G=1$.
In addition to the ansatz~\eqref{ansatz} we consider the metric form
\begin{equation}
\label{metric}
 ds^2=-e^{-2\delta(r)}N(r) dt^2+\frac{dr^2}{N(r)}+r^2 (d\theta^2+\sin^2\theta d\varphi^2)\ , \qquad {\rm where} \ \ \
N(r)\equiv 1-\frac{2m(r)}{r} \ ,
\end{equation}
and $m(r)$ is the Misner-Sharp mass; $r$ is thus the areal radius, a geometrically meaningful coordinate.

The ansatz~\eqref{ansatz} and~\eqref{metric} yield the following effective Lagrangian:
\begin{eqnarray}
\label{Leff}
\mathcal{L}_{\rm eff}=e^{-\delta}m'-\frac{r^2}{2}e^{-\delta} N \phi'^2+\frac{r^2 }{2}  f(\phi)e^{\delta} V'^2 \ .
\end{eqnarray}
As in the flat spacetime case, the equation of the electric potential possesses a first integral, which generalises~\eqref{maxint}, and reads
\begin{eqnarray}
\label{first-int}
V'= e^{-\delta } \frac{Q}{r^2 f(\phi)} \ ,
\end{eqnarray}
where again the integration constant $Q$ is the electric charge, which we shall assume to be strictly positive,
without any loss of generality. Using this integral, the remaining equations of motion become\footnote{There is an extra equation, which is a constraint and can be derived from 
(\ref{ec-m})-(\ref{ec-phi}). }
\begin{eqnarray}
\label{ec-m}
&&
m'=\frac{r^2}{2}  N \phi'^2+\frac{Q^2}{2r^2} S(\phi) \ ,
\\
\label{ec-d}
&&
\delta'+r \phi'^2=0 \ , 
\\
&&
\label{ec-phi}
(e^{-\delta} r^2 N\phi')'-\frac{ e^{-\delta}}{2r^2  }  \frac{dS(\phi)}{d\phi} Q^2 =0\ ,
\end{eqnarray}
where we have defined
\begin{eqnarray} 
S(\phi)\equiv \frac{1}{f(\phi)}\ .
\end{eqnarray}

The smooth of a spacetime configurations can be assessed by analysing the Ricci scalar
\begin{eqnarray}
\label{Ricci}
 R=\frac{N}{r}(3r\delta'-4)+\frac{2}{r^2}
\left[
1+N(r^2\delta''-(1-r\delta')^2)
\right]
-N'' \ ,
\end{eqnarray}
and the Kretschmann scalar
\begin{eqnarray}
\label{Kr}
K=\frac{4}{r^4}(1-N)^2
+\frac{2}{r^2}
\left[
N'^2+(N'-2N\delta')^2
\right]
+
\left[
N''-3\delta'N'+2N(\delta'^2-\delta'')
\right]^2.
\end{eqnarray}

\subsection{Asymptotic expansions}
\label{sec31}

\subsubsection{Near the origin}
A small $r$ analysis of the field equations confirms the conclusion observed in the flat space analysis: for a scalar field admitting a power series expansion near the origin $\phi=\phi_0+\phi_1 r+\dots$ and $\phi_1\neq 0$, if the coupling diverges as $1/r^4$, finite energy, everywhere regular solutions are possible.  To see this, we again start by assuming the existence of a power series expansion of solutions,
with
the scalar field approaching a finite nonzero value
\begin{eqnarray} 
 \phi(r) \to \phi_0~~{\rm as}~~ r\to 0 \ ,
\end{eqnarray}
where $\phi_0$ is arbitrary. Then, 
the equations of motion, together with the assumption of regularity, impose, for the $n^{th}$ derivative of $S(\phi)$ computed at the origin, denoted $S^{(n)}(\phi_0)$,
\begin{eqnarray} 
S(\phi_0)=S^{(1)}(\phi_0)=S^{(2)}(\phi_0)=S^{(3)}(\phi_0)=0 \ , \qquad {\rm whereas} \qquad S^{(4)}(\phi_0)>0 \ .
\end{eqnarray}
This implies the advertised behaviour: the coupling function $f(\phi)$ diverges as $1/r^4$ as $r\to 0$.
This behaviour cancels the divergence  associated with the presence of an electric charge, providing a smooth configuration as $r\to 0$.

The small $r$ expansion of the matter functions reads
\begin{eqnarray} 
\phi(r)=\phi_0-\frac{2\sqrt{6} r}{Q \sqrt{S^{(4)}(\phi_0)}} +\phi_2 r^2+\dots \ , \qquad
V(r)=-\frac{8 e^{-\delta_0}}{Q^3 S^{(4)}(\phi_0)}r^3+\dots \ ,
\end{eqnarray}
while for the metric functions we find
\begin{eqnarray} 
m(r)=\frac{8}{Q^2}\frac{1}{S^{(4)}(\phi_0)}r^3
-\frac{2\sqrt{6} \phi_2}{Q \sqrt{S^{(4)}(\phi_0)}}r^4+\dots \ , \qquad 
\delta(r)=\delta_0-\frac{12 r^2}{Q^2 S^{(4)}(\phi_0)}+\dots \ ,
\end{eqnarray}
where $\phi_2$ and $\delta_0$ are constants that are fixed by the numerics when integrating the field equations from the origin to infinity and requiring the correct asymptotic behaviour.  With this expansion,
both the  Kretschmann curvature scalar and Ricci scalar are finite as $r\to 0$, taking the form
\begin{equation} 
K\equiv R_{\mu\nu\alpha\beta}R^{\mu\nu\alpha\beta} =\frac{3840}{Q^4 [S^{(4)}(\phi_0)]^2}-\frac{2560 \sqrt{6} \phi_2  }{Q^3 [S^{(4)}(\phi_0)]^{3/2}}r+\dots \ ,
\end{equation}
and
\begin{equation}
R=\frac{48}{Q^2 S^{(4)}(\phi_0)}
-\frac{16 \sqrt{6} \phi_2  }{Q  (S^{(4)}(\phi_0))^{1/2}}r+\dots \ . 
\end{equation}

The small $r$ expansion of $S(\phi)$ reads
\begin{eqnarray} 
 S(\phi)=\frac{24}{Q^4 S^{(4)}(\phi_0)}r^4-\frac{8\sqrt{6}\phi_2}{Q^3\sqrt{S^{(4)}(\phi_0)}}r^5+\dots\ ,
\end{eqnarray}
which implies the following $generic$ approximate form of the coupling function
\begin{eqnarray} 
\label{condition}
S(\phi) =\frac{1}{f(\phi)} \sim (\phi-\phi_0)^4 ~~~{\rm as}~~r\to 0
 \ .
\end{eqnarray}
Of course, we could have assumed that in the scalar field expansion $\phi_1=0$ and the power series starts at a  higher order term. This would impact in the way the coupling diverges at the origin, similarly to the flat spacetime analysis of section~\ref{regsol}. For concreteness, here we focus on the case with $\phi_1\neq 0$.

\subsubsection{Near infinity}

A large $r$ analysis of the field equations, on the other hand,  imposing
\begin{eqnarray}
\label{cond-inf}
f(\phi) \to 1~~{\rm as}~~r\to \infty,
\end{eqnarray}
yields the following approximate solutions:
\begin{equation}
\label{inf1}
m(r)=M-\frac{Q^2 +Q_s^2}{2r}+\dots \ , \qquad
\phi(r)=\frac{Q_s}{r}+\dots \ ,
\end{equation}
\begin{equation}
V(r)=\Phi -\frac{Q}{r}+\dots\ , \qquad
\delta(r)=\frac{Q_s^2}{2r^2}+\dots \ .
\end{equation}
Here $M$ is the ADM mass and $Q$ is the electric charge; 
$\Phi $ is the electrostatic potential at infinity  and $Q_s$ is the scalar 'charge' which in general needs not equal the electric charge (it \textit{did} in the flat spacetime illustration above). In fact, the equations of motion possess again two first integrals implying  that the $gravitating$
solutions satisfy the following relation
\begin{eqnarray}
\label{figen}
 M^2+Q_s^2=Q^2 .
\end{eqnarray} 
This last equation, in particular, shows manifestly the curved background breaks the equality between $Q$ and $Q_s$.

Interestingly, one can show that there is a Smarr relation in terms of these asymptotic quantities, which is not affected by the scalar field,
\begin{eqnarray}
\label{Smarr}
M= \Phi  Q \ . 
\end{eqnarray}
Moreover, a first law of thermodynamics can be obtained in the form 
\begin{eqnarray}
\label{1st}
dM= \Phi  dQ \  .
\end{eqnarray}
We emphasise the absence of a scalar field contribution in these relations.

\subsection{The full solutions}
\label{sec32}

The gravitating version of the exact solution in Minkowski spacetime described in subsection~\ref{sec241}, with coupling~\eqref{coupling1}, and whose asymptotic limits have been described in subsection~\ref{sec31}, can be constructed numerically.  The set of four ordinary differential equations obtained from the above setup was solved numerically by using a standard Runge-Kutta ordinary differential
equation solver and appropriate boundary conditions. Fixing $\alpha$, gravitating solitons exist for arbitrary large values of $Q$. The profile of a typical solution is shown in Fig.~1.
 \begin{figure}[h!]
\begin{center}
\includegraphics[width=0.55\textwidth]{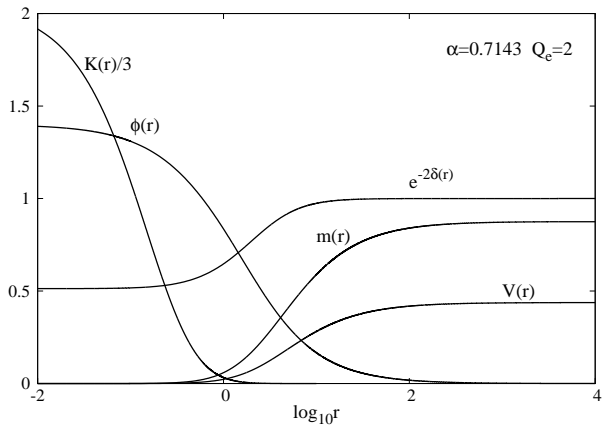}
\caption{\small{ 
Profiles of an illustrative gravitating soliton with the coupling~\eqref{coupling1}.
}}
\label{profile1}
\end{center}
\end{figure} 
As one can see, the profiles of the various functions, and in particular that of the Kretschmann scalar $K$, are smooth as $r\to 0$. 
In Fig.~\ref{profile2} we show the ADM mass $vs.$ electric charge diagram for families of solutions with different values of $\alpha$. One can see that for all families, the solutions trivialise as $\alpha\rightarrow 0$. Moreover, a smaller $\alpha$ implies that the same value of the electric charge can support a more massive soliton. Obviously, the solutions also trivialise as $Q\rightarrow 0$. The electric charge supports the soliton. This is also manifest from the following  virial identity that can be derived for these solutions:
\begin{eqnarray}
\label{virial} 
\int_{0}^\infty dr  ~
e^{-\delta}  \phi'^2
=
\int_{0}^\infty dr  ~
 \frac{e^{-\delta } }{r^2} 
\frac{Q^2}{f(\phi)}  \ .
\end{eqnarray} 
For $Q=0$ the right hand side vanishes, and so must the left hand side, which implies $\phi'=0$ and hence no non-trivial scalar profile exists.

 \begin{figure}[h!]
\begin{center}
\includegraphics[width=0.55\textwidth]{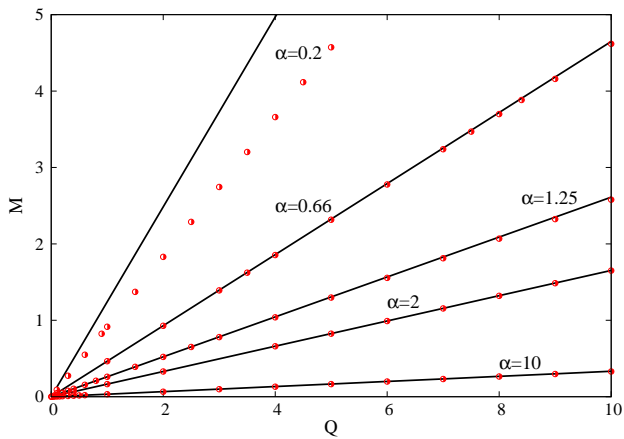}  
\caption{\small{ 
ADM $vs.$ electric charge  for families of gravitating solitons with different values of $\alpha$. The straight lines are obtained from the perturbative solutions, whereas the dots represent the numerical solutions. 
}}
\label{profile2}
\end{center}
\end{figure} 

Self-gravitating solitons with the coupling~\eqref{cfgen} and $n=3$ where also obtained. They follow the same pattern as the $n=3$ case, which is therefore illustrative.

\subsection{Perturbative solutions}
\label{sec33}
The existence of a flat spacetime solution, whose total mass-energy is proportional to $1/\alpha$, suggests that the self-gravitating solitons may be expressed as a perturbative series in $1/\alpha$. Let us indeed show that the numerical solutions of the previous subsection can be approximated by such perturbative solutions. This approximation, as we will show and as one may anticipate, is accurate for sufficiently large $\alpha$.

The perturbative solutions are obtained by performing a power series expansion for all relevant functions
\begin{eqnarray}
&& m(r)=\sum_{k\geqslant 1}\left(\frac{1}{\alpha}\right)^{k} m_k(r) \ , \qquad \delta(r)=\sum_{k\geqslant 1} \left(\frac{1}{\alpha}\right)^{k}\delta_k(r) \ , \\
&& \phi(r)=\sum_{k\geqslant 1} \left(\frac{1}{\alpha}\right)^{k}\phi_k(r) \ , \qquad
  V(r)=\sum_{k\geqslant 1} \left(\frac{1}{\alpha}\right)^{k} V_k(r) \  .
\end{eqnarray}

As for the numerical solutions of the previous subsection, we focus on the quartic coupling function~\eqref{coupling1}. Solving iteratively the field equations order by order in $1/{\alpha}$,
we arrive at the following expressions\footnote{We have computed the solution up to eighth order and no obvious pattern could be found. Here we display only the first few terms for each function.}
\begin{eqnarray}
&&
m_{1}(r)=0=m_3(r) \ , \qquad
m_{2}(r)=\frac{q r^3}{3(q+r)^3}\ , \qquad m_4(r)=-\frac{ q r^4(10 q^2+q r+r^2)}{90 (q+r)^6} \ , \nonumber
\\
&&
\nonumber
\delta_{1}(r)=0=\delta_3(r) \ ,\qquad
\delta_2(r)=\frac{q^2(q+3r)}{6(q+r)^3} \ , \nonumber \\
&&
\delta_4(r)=\frac{q^2(q^4+6 q^3 r+15 q^2 r^2+100 q r^3-30r^4)}{540 (q+r)^6} \ ,
\\
&&
\nonumber
\phi_1(r)=\frac{q}{q+r}\ , \qquad \phi_2=0 \ , \qquad \phi_3(r)=\frac{q (2 q-r)r^2}{18 (q+r)^4} \ ,
\\
&&
\nonumber
V_1(r)=\frac{r^3}{3(q+r)^3}\ , \qquad V_2(r)=0\ , \qquad 
V_3(r)=-\frac{r^3(5 q^3+25 q^2 r+6 q r^2+r^3)}{90 (q+r)^6} \ ,
\end{eqnarray}
where $q$ is a free parameter, whose physical significance becomes
transparent by computing the far field asymptotics of the electric potential.
One finds it is related to the electric charge measured at infinity $Q$, as 
\begin{eqnarray}
Q=\frac{q}{\alpha} \ .
\end{eqnarray}

The perturbative solution yields the following ADM mass and scalar charge, valid to fourth order in perturbation theory:
\begin{eqnarray}
M=\frac{Q}{3 \alpha}   \left(1-\frac{1}{30 \alpha^2}+\frac{1}{1080 \alpha^4}\right) \ , \qquad 
Q_s= \left(1-\frac{1}{18 \alpha^2}+\frac{7 }{3240 \alpha^4}\right)Q \ . 
\label{mqper}
\end{eqnarray}
Observe that the first terms in~\eqref{mqper} reproduce the flat spacetime limit, eq.~\eqref{totale1} and the fact that the electric and scalar charge coincide in that limit.

In Fig.~\ref{profile2} the perturbative solutions~\eqref{mqper} are compared with the numerical solutions. One can observe that the former provide a good approximation for large values of 
$\alpha$; for instance, for $\alpha=10$ the relative difference between the numerical result 
for $M(Q)$  and the theory one is around $10^{-4}$.
However, the differences start  to increase for smaller $\alpha$. This is illustrated by  the results for $\alpha=0.2$ in Fig.~\ref{profile2}.

Finally let us mention that a similar solution has been  derived for the self-gravitating solitons with the coupling~\eqref{cfgen} and $n=3$. In this case one finds, $e.g.$ 
\beq
 M=\frac{2Q}{5 \alpha}   \left(1-\frac{2}{45 \alpha^2}+\frac{22}{14625 \alpha^4}\right)+\dots \ .
\eeq

\section{Discussion}
\label{sec4}
Recently, a set of theorems were shown~\cite{Herdeiro:2019oqp} establishing that the model~\eqref{actiontotal} does not allow self-gravitating solitons. One of the observations therein is that if the coupling would diverge the theorems could, potentially, be circumvented. The purpose of this paper was to provide the mechanism how this can happen providing a simple construction of flat spacetime and gravitating solitons.

Preliminary analysis shows the solitons we have described herein are stable against spherical perturbations. If this is the case for generic perturbations, these solitons can be used for dynamical studies in many setups, as, for instance, boson stars~\cite{Liebling:2012fv}. Moreover, this construction reveals how to de-singularise the Coulomb field in a classical field theory, without resorting to non-linear electrodynamics, as in the Born-Infeld model~\cite{Born:1934gh}, or invoking a manifestly extended object, such as in the Dirac model of the electron as a spherical membrane~\cite{Dirac:1962iy}.

Finally, let us remark that there is a well known similarity between the EMS model and the extended scalar-tensor-Gauss-Bonnet model, where black hole scalarisation was first pointed out in~\cite{Doneva:2017bvd, Silva:2017uqg,Antoniou:2017acq}. Very recently, a family of particle-like solutions in the latter model were discussed~\cite{Kleihaus:2019rbg}. These particle like solutions are also supported by a divergent coupling making them the counterparts of the solutions described herein. But in the cases reported in~\cite{Kleihaus:2019rbg} the scalar field also diverges at the origin, in contrast with our fully regular solutions.

\section*{Acknowledgements}
This work is supported by the Funda\c{c}\~ao para a Ci\^encia e a Tecnologia (FCT) project UID/MAT/04106/2019 (CIDMA) and by national funds (OE), through FCT, I.P., in the scope of the framework contract foreseen in the numbers 4, 5 and 6
of the article 23, of the Decree-Law 57/2016, of August 29,
changed by Law 57/2017, of July 19. We acknowledge support  from the project PTDC/FIS-OUT/28407/2017.   This work has further been supported by  the  European  Union's  Horizon  2020  research  and  innovation  (RISE) programmes H2020-MSCA-RISE-2015
Grant No.~StronGrHEP-690904 and H2020-MSCA-RISE-2017 Grant No.~FunFiCO-777740. The authors would like to acknowledge networking support by the
COST Action CA16104.




\begin{thebibliography}{99}
\bibitem{Herdeiro:2019oqp}
  C.~A.~R.~Herdeiro and J.~M.~S.~Oliveira,
  Class.\ Quant.\ Grav.\  {\bf 36} (2019) no.10,  105015
  [arXiv:1902.07721 [gr-qc]].
\bibitem{Kaluza:1921tu}
  T.~Kaluza,
  Sitzungsber.\ Preuss.\ Akad.\ Wiss.\ Berlin (Math.\ Phys.\ ) {\bf 1921} (1921) 966
   [Int.\ J.\ Mod.\ Phys.\ D {\bf 27} (2018) no.14,  1870001]
  [arXiv:1803.08616 [physics.hist-ph]].
\bibitem{Klein:1926tv}
  O.~Klein,
  Z.\ Phys.\  {\bf 37} (1926) 895
   [Surveys High Energ.\ Phys.\  {\bf 5} (1986) 241].
\bibitem{Appelquist:1987nr} 
  T.~Appelquist, A.~Chodos and P.~G.~O.~Freund,
  Reading, USA: Addison-Wesley (1987) 619 (Frontiers in physics, 65).
\bibitem{VanNieuwenhuizen:1981ae}
  P.~Van Nieuwenhuizen,
  Phys.\ Rept.\  {\bf 68} (1981) 189.
\bibitem{Martin:2007ue}
  J.~Martin and J.~Yokoyama,
  JCAP {\bf 0801} (2008) 025
  [arXiv:0711.4307 [astro-ph]].
\bibitem{Maleknejad:2012fw}
  A.~Maleknejad, M.~M.~Sheikh-Jabbari and J.~Soda,
  Phys.\ Rept.\  {\bf 528} (2013) 161
  [arXiv:1212.2921 [hep-th]].
\bibitem{Herdeiro:2018wub}
  C.~A.~R.~Herdeiro, E.~Radu, N.~Sanchis-Gual and J.~A.~Font,
  Phys.\ Rev.\ Lett.\  {\bf 121} (2018) no.10,  101102
  [arXiv:1806.05190 [gr-qc]].
\bibitem{Fernandes:2019rez}
  P.~G.~S.~Fernandes, C.~A.~R.~Herdeiro, A.~M.~Pombo, E.~Radu and N.~Sanchis-Gual,
  Class.\ Quant.\ Grav.\  {\bf 36} (2019) no.13,  134002
  [arXiv:1902.05079 [gr-qc]].
\bibitem{Astefanesei:2019pfq}
  D.~Astefanesei, C.~Herdeiro, A.~Pombo and E.~Radu,
  JHEP \ {\bf 10} (2019)  78.
  [arXiv:1905.08304 [hep-th]].
\bibitem{Myung:2018vug}
  Y.~S.~Myung and D.~C.~Zou,
  Eur.\ Phys.\ J.\ C {\bf 79} (2019) no.3,  273
  [arXiv:1808.02609 [gr-qc]].
\bibitem{Myung:2018jvi}
  Y.~S.~Myung and D.~C.~Zou,
  Phys.\ Lett.\ B {\bf 790} (2019) 400
  [arXiv:1812.03604 [gr-qc]].
\bibitem{Myung:2019oua}
  Y.~S.~Myung and D.~C.~Zou,
  Eur.\ Phys.\ J.\ C {\bf 79} (2019) no.8,  641
  [arXiv:1904.09864 [gr-qc]].
\bibitem{Konoplya:2019goy}
  R.~A.~Konoplya and A.~Zhidenko,
  Phys.\ Rev.\ D {\bf 100} (2019) no.4,  044015
  [arXiv:1907.05551 [gr-qc]].
\bibitem{Fernandes:2019kmh}
  P.~G.~S.~Fernandes, C.~A.~R.~Herdeiro, A.~M.~Pombo, E.~Radu and N.~Sanchis-Gual,
  arXiv:1908.00037 [gr-qc].

\bibitem{Boskovic:2018lkj}
  M.~Boskovic, R.~Brito, V.~Cardoso, T.~Ikeda and H.~Witek,
  arXiv:1811.04945 [gr-qc].
\bibitem{Minamitsuji:2018xde}
  M.~Minamitsuji and T.~Ikeda,
  arXiv:1812.03551 [gr-qc].
\bibitem{Silva:2018qhn}
  H.~O.~Silva, C.~F.~.B.~Macedo, T.~P.~Sotiriou, L.~Gualtieri, J.~Sakstein and E.~Berti,
  arXiv:1812.05590 [gr-qc].
\bibitem{Herdeiro:2019yjy}
  C.~A.~R.~Herdeiro and E.~Radu,
  Phys.\ Rev.\ D {\bf 99} (2019) no.8,  084039
  [arXiv:1901.02953 [gr-qc]].




\bibitem{Peskin:1995ev}
  M.~E.~Peskin and D.~V.~Schroeder,
  ``An Introduction to quantum field theory,''
  Taylor \& Francis Inc., 1995.

\bibitem{Derrick:1964ww}
  G.~H.~Derrick,
  J.\ Math.\ Phys.\  {\bf 5} (1964) 1252.


\bibitem{Gibbons:1990um}
  G.~W.~Gibbons,
  Lect.\ Notes Phys.\  {\bf 383} (1991) 110
  [arXiv:1109.3538 [gr-qc]].


  
\bibitem{Liebling:2012fv}
  S.~L.~Liebling and C.~Palenzuela,
  Living Rev.\ Rel.\  {\bf 15} (2012) 6
   [Living Rev.\ Rel.\  {\bf 20} (2017) no.1,  5]
  [arXiv:1202.5809 [gr-qc]].

\bibitem{Born:1934gh}
  M.~Born and L.~Infeld,
  Proc.\ Roy.\ Soc.\ Lond.\ A {\bf 144} (1934) no.852,  425.

\bibitem{Dirac:1962iy}
  P.~A.~M.~Dirac,
  Proc.\ Roy.\ Soc.\ Lond.\ A {\bf 268} (1962) 57.


\bibitem{Doneva:2017bvd}
  D.~D.~Doneva and S.~S.~Yazadjiev,
  Phys.\ Rev.\ Lett.\  {\bf 120} (2018) no.13,  131103
  [arXiv:1711.01187 [gr-qc]].
\bibitem{Silva:2017uqg}
  H.~O.~Silva, J.~Sakstein, L.~Gualtieri, T.~P.~Sotiriou and E.~Berti,
  Phys.\ Rev.\ Lett.\  {\bf 120} (2018) no.13,  131104
  [arXiv:1711.02080 [gr-qc]].
\bibitem{Antoniou:2017acq}
  G.~Antoniou, A.~Bakopoulos and P.~Kanti,
  Phys.\ Rev.\ Lett.\  {\bf 120} (2018) no.13,  131102
  [arXiv:1711.03390 [hep-th]].
\bibitem{Kleihaus:2019rbg}
  B.~Kleihaus, J.~Kunz and P.~Kanti,
  arXiv:1910.02121 [gr-qc].


\end{thebibliography}
\end{document}